\documentclass{rspublic}

\usepackage{graphicx}
\usepackage{times}
\usepackage{color}
\usepackage{latexsym,wasysym}
\usepackage{amssymb}

\newcommand{\be}{\begin{equation}}
\newcommand{\ee}{\end{equation}}

\newcommand{\ket}[1]{\left | #1\right \rangle}
\newcommand{\bra}[1]{\left \langle #1\right |}

\begin{document}

\title[Why should anyone care about computing with anyons?]
{Why should anyone care about computing with anyons?}

\author[Gavin K. Brennen and Jiannis K. Pachos]
{Gavin K. Brennen$^1$ and Jiannis K. Pachos$^2$}

\affiliation{
$^1$Institute for Quantum Optics and Quantum Information, Techniker Str.
21a, 6020, Innsbruck, Austria\\
$^2$School of Physics and Astronomy, University of Leeds, Leeds LS2 9JT, UK}

\label{firstpage}

\maketitle

\begin{abstract}{Topological Quantum Computation, Topological models.}

In this article we present a pedagogical introduction of the main ideas and
recent advances in the area of topological quantum computation. We
give an overview of the concept of anyons and their exotic statistics,
present various models that exhibit topological behavior, and we establish
their relation to quantum computation. Possible directions for the physical
realization of topological systems and the detection of anyonic behavior are
elaborated.

\end{abstract}

\section{Introduction}

The objectives of current research in quantum computation are mainly
twofold. Firstly, to perform neat quantum evolutions, robust against
decoherence and control errors. Secondly, to find new quantum algorithms
that outperform their classical counterparts. Topological quantum systems
have been proven to be a fertile environment for addressing both of these
questions. In particular, topological quantum computation is concerned with
two dimensional many body systems that support excitations with exotic
statistics called anyons. Encoding and manipulating information with anyons
is an intrinsically error free procedure. If physically realized it could
allow reliable quantum computation without the need of the huge error
correction overhead. Moreover, it has provided the setup for constructing a
new quantum algorithm inspired from the evolution of topological
excitations. This algorithm can approximated the Jones polynomials,
topological invariants that can distinguish between inequivalent knots. The
exact evaluation of the Jones polynomials is an exponentially hard classical
problem with much interest in various fields such as in biology, through the
protein folding problem, and in statistical physics.

Topological quantum systems are more familiar from the Quantum Hall Effect
where a two dimensional layer of electrons is subject to a strong vertical
magnetic field. The low energy spectrum of these systems is governed by a
trivial Hamiltonian, $H=0$. Nevertheless, they have an interesting behavior
due to the non-trivial statistics of their excitations. It has been proven
that this behavior is dictated by the presence of anyons. Unlike bosons or
fermions, anyons have a non-trivial evolution when one circulates another.
Lately, lattice counterparts to these continuous systems have been proposed
that can also support anyonic excitations. These spin or electron systems
offer an alternative way of generating and manipulating anyons that can also
support topological quantum computation.

Here we will give a pedagogical presentation of various inspiring ideas
related to topological quantum computation that manifests the close relation
between physics, mathematics and information theory. We will present several
topological systems, explain their relevance to error free quantum
computation and give possible physical realizations with cold atomic systems.
 The interested reader may refer to the bibliography for the in
depth analysis of the subjects presented here.

\section{Anyons: what are they?}

\begin{figure}
\begin{center}
\includegraphics[width=1.2in]{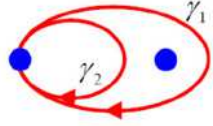}
\caption{A particle spans a loop around another one. In three dimensions it
is possible to continuously deform the path $\gamma_1$ to the path
$\gamma_2$ which is equivalent to a trivial path.}
\label{fig:anyons1}
\end{center}
\end{figure}

It is commonly accepted that point-like particles, elementary or not, come
in two species:  bosons or fermions.  The particle label is determined by
 the behavior of their wave function: if two identical non-interacting particles are
exchanged we expect that their wavefunction will acquire either a plus sign
(bosons) or a minus sign (fermions).  These are the only observed
statistical behaviors for particles that exist in our three dimensional
world. Indeed, circulating a particle around an identical one spans a path
that in three dimensions can be continuously deformed to a trivial path as
seen in Figure~\ref{fig:anyons1}.
As a consequence the wave function, $\ket{\Psi(\gamma_1)}$, of the system
after the circulation has to be exactly the same as the original one
$\ket{\Psi(0)}$, i.e.
\be
\ket{\Psi(\gamma_1)} = \ket{\Psi(\gamma_2)} = \ket{\Psi(0)}
\label{eqn:bosefermi1}
\ee
It is easily seen that a full circulation is equivalent to two successive
particle exchanges. Thus, a single exchange can result to a phase factor
$e^{i\varphi}$ that has to square to unity in order to be consistent with
(\ref{eqn:bosefermi1}), giving, finally, $\varphi = 0,\pi$. These two cases
correspond to bosonic and fermionic statistics respectively.
\footnote{More precisely, they correspond to one dimensional
irreducible representations of the symmetric group. Higher dimensional
representations, with parastastics, could also occur but are not observed in
nature.  Anyons are classified by irreducible representations of the braid
group.}

\begin{figure}
\begin{center}
\includegraphics[width=4in]{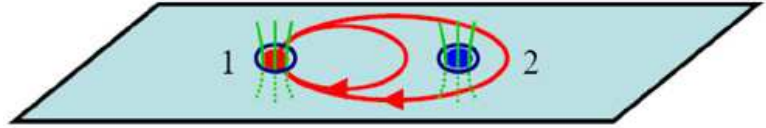}
\caption{In two dimensions the two paths $\gamma_1$ and $\gamma_2$ are
topologically distinct. This gives the possibility of having non-trivial
phase factors appearing when one particle circulates the other. This can be
visualized by having the particles carrying charge as well as magnetic flux
giving rise to the Aharonov-Bohm effect.}
\label{fig:anyons2}
\end{center}
\end{figure}

When we restrict to two spatial dimensions there are more possibilities in
the statistical behavior of the particles. If the particle circulation
$\gamma_1$ of Figure~\ref{fig:anyons1} is performed on a plane, then it is
not possible to continuously deform it to the path $\gamma_2$. Still the
evolution that corresponds to $\gamma_2$ is equivalent to the trivial
evolution as seen in Figure~\ref{fig:anyons2}. As we are not able to deform
the evolution of path $\gamma_1$ to the trivial one, it is possible to
assign an arbitrary phase factor or even a whole unitary to this evolution.
Thus, particles in two dimensions can have richer statistical behavior
different from bosons or fermions.

To visualize the behavior of anyons one can think of them as being composite
particles consisting of a flux $\Phi$ and a ring of charge $q$ as depicted
in Figure \ref{fig:anyons2}.
If the particle $1$ circulates particle $2$ then its charge $q$ goes around
the flux $\Phi$ thus acquiring a phase factor $U=e^{iq\Phi}$ due to the
Aharonov-Bohm effect. The statistical angle of these anyons is then
$\varphi=q\Phi/2$. Non-abelian charges and fluxes can generate unitary
matrices instead of phase factors. In this case a circulation of one anyon
around another can lead to a final state in superposition. The anyons that have
such statistics are called non-abelian, while anyons that obtain a simple
phase factor are called abelian. Note that the generation of the evolution
is not a result of a direct interaction but rather a topological consequence
as in the Aharonov-Bohm effect. In reality the presence of charge and flux
come from an effective gauge theory that describes the low energy behavior
of the model.

\begin{figure}
\begin{center}
\includegraphics[scale=0.4]{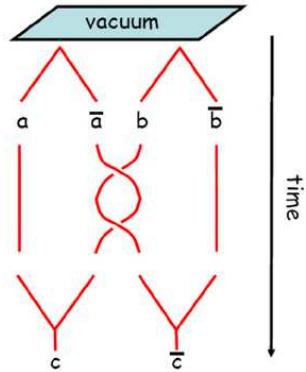}
\caption{The world lines of the anyons where the third dimension
depicts time running downwards. From the vacuum two pairs of anyon and
anti-anyon are generated depicted by $(a,\bar a)$ and $(b,\bar b)$. Then
anyons $\bar a$ and $b$ are braided by circulating one around the other.
Finally, the anyons are pairwise fused, but they do not necessarily return
to the vacuum as the braiding process may have changed their internal
state.}
\label{fig:anyons3}
\end{center}
\end{figure}

As the statistical properties dominate the behavior of the anyonic states,
it is convenient to employ the world lines of the particles to keep track of
their positions. In this way exchanges of the anyons can be easily describe
just by braiding their world lines. Moreover, we can depict the pair
creation of anyon anti-anyon from the vacuum as well as their annihilation
(see Figure~\ref{fig:anyons3}).
Out of the vacuum an anyon and an anti-anion are generated with a local
physical process. We assume that we can trap and move the anyons around the
plane leading to world lines in $2+1$ dimensions. In the particular depicted
example two pairs of anyons are created $(a,\bar a)$ and $(b,\bar b)$. Then
anyons $\bar a$ and $b$ are braided by circulating one around the other and
then the corresponding pairs are fused. The fusion may not result to the
vacuum as the braiding process could change the internal state of one of the
anyons. The outcomes of the fusions are the anyons $c$ and $\bar c$. They
can be further fused giving the vacuum that we had started with in agreement
with the conservation of the relevant quantum numbers.

As we have seen the generation of the anyons with pair creation results in
well defined pair of anyon anti-anyon. When two anyons are fused, it is
possible to have various outcomes depending on their internal state. In
Figure~\ref{fig:anyons3}, one can see that the fusions result in the anyons
$c$ and $\bar c$. In general, depending in the particular internal state of
the anyons, one can have different unique outcomes from the fusion. This is
denoted in the following way
\begin{equation}
a\times b = N^c_{ab} c+N^d_{ab} d+...
\end{equation}
where anyons $a$ and $b$ are fused to produce either anyon $c$ or $d$ or any
other possible outcome. The order of the outcomes in the sum is not
important. The integers $N_{jk}^l$ denote the multiplicity that the
particles $l$ are generated out of the fusion of the particles $j$ and $k$.
This is similar to the tensor product of spins that results in a new spin
basis, e.g. $ {1\over 2}\otimes {1\over 2} = 0\oplus 1 $. In particular
abelian anyons have only a single fusion product $a\times b = c$, while for
non-abelian anyons the outcome of the fusion is not unique. In particular,
one can assign a unitary matrix $F$ that transforms between the different
ways that three anyons can fuse to give a fourth one.

\begin{figure}
\begin{center}
\includegraphics[scale=0.5]{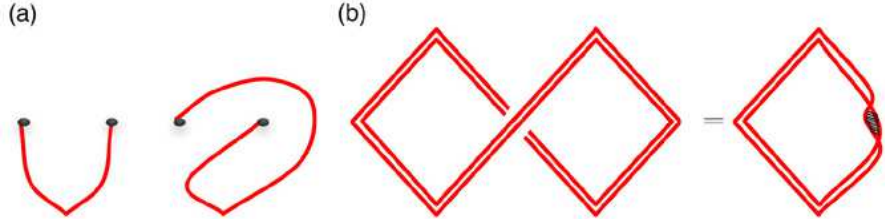}
\caption{(a) Two anyons are exchanged with paths attached to them from a common
reference point. The exchange is described by a unitary matrix $R$ acting on
the state of the anyons. (b) The equivalence between spin and statistics.
Two anyons, each one from two anyon anti-anyon pairs, are exchanged and
recombined fusing to the vacuum. This process is equivalent to rotating one
anyon by $2\pi$ from a pair and recombining them causing an evolution due to
the spin of the anyon. The anyons are depicted here as ribbons to keep track
of their spin rotation.}
\label{fig:statistics}
\end{center}
\end{figure}

Apart from the fusion rules we are also interested in the statistics of the
anyons. The latter is characterized by the unitary $R$ that describes the
evolution of the anyonic states when the anyons are anticlockwise
interchanged (see Figure \ref{fig:statistics}(a)).
The interpretation of the anyons with world lines, and in particular with
``world ribbons" makes apparent the connection between statistics and spin.
Indeed, in Figure \ref{fig:statistics} we see the schematical equivalence
between the process of exchanging two abelian anyons and the rotation of an
anyon by $2\pi$. The first is evolved by the statistical unitary $R$ (a
phase factor for abelian anyons) and the second is expected to obtain a
phase factor $e^{i2\pi J}$, where $J$ is the spin of the anyon. A direct
application leads to the connection between the integer spins for bosons and
half integer spins for fermions. This is in agreement with the
interpretation of anyons as a composite object of charge and flux. In this
case the rotation of an anyon by $2\pi$ will rotate the charge around the
flux leading to the same phase factor dictated by their statistics due to
the Aharonov-Bohm effect.

Finally, one can show that consistency equations can be considered that give
a relation between the statistical processes and the fusion relations. Two
such basic processes, the braiding of two anyons and the reordering of the
fusion of three anyons are given by the unitaries $R$ and $F$, respectively.
These consistency equations are called pentagon and hexagon equations ( Turaev 1994).

\section{Surface codes}
\label{surf}

The marriage of topology and quantum information began with a proposal
(Kitaev 1997) for encoding quantum information in the collective state of
interacting spins on a surface. Such codes are known as surface codes and
have several desirable characteristics. First the information is encoded in
a subspace of the state-space of the spins, whose dimension is a property of
the topology of the surface alone. This is a global  property robust to
local perturbations. Second, this codespace corresponds to the degenerate
ground subspace of a Hamiltonian that is $\emph{quasi-local}$. Here
quasi-local means that the interactions take place only among a few
neighboring particles (typically $3-6$).  This is in stark contrast with
most other quantum error correction codes which do not correspond to the
ground state of any quasi-local Hamiltonian. Also, the ground states possess
a symmetry. They are invariant under the action of a product of
connected local spin operators which form a closed loop, while the action
of an open string of operators creates excitations at its endpoints. The
ground states are isolated from low lying excited states by an energy gap
which remains finite in the thermodynamic limit. Provided the temperature of
the environment is much less than this gap, the probability of errors
(excitations) is small. Third, any local errors that do arise are
characterized as anyonic quasiparticle pairs, whose mass is proportional to
the energy gap. These anyons are positioned at the ends of the open string
operations acting on the vacuum (ground states). All that is required to
correct the errors is to annihilate the anyons by connecting the string ends
along a closed loop that is homologically trivial, i.e. it can be contracted
to a point. Logical operations on the codespace correspond to loops of
operations that are topologically non-trivial, i.e. they cannot be shrunk to
a point loop.

\begin{figure}
\begin{center}
\includegraphics[width=\columnwidth]{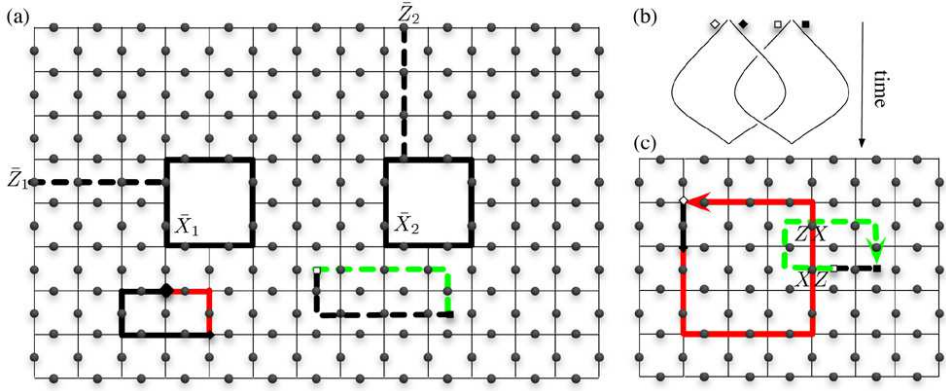}
\caption{\label{fig:surf}Surface codes and anyonic excitations. (a) A twice punctured plane
encoding $2$ qubits. Physical qubits (grey dots) reside on the edges and a
black solid (dashed) string intersecting a qubit corresponds to a physical
$X(Z)$ operation on that qubit. The logical operations $\bar{X}_{1,2}$
correspond to non-trivial loops around the punctures while the logical
$\bar{Z}_{1,2}$ connect boundaries. This code protects against $3$ bit flip
errors and $2$ phase errors but a more resilient code results by increasing
the lattice size. Correctible errors are shown in the lower half of the
surface.  String boundaries, denoted by $\square$ or
$\diamond$ indicate where a vertex or face stabilizer condition is violated.
These errors can be corrected by connecting the string ends along a
contractible loop (red and green strings). (b)  The action of braiding a
charge $(\diamond)$ around a flux $(\square)$.  For a $\mathbb{Z}_2$ gauge
theory, particles and anti-particles are equivalent. (c)  Configuration path
for the braiding beginning in the vacuum, i.e. a ground state.  The action
on the state is obtained by interchanging the order of operations on one
intersecting edge.  Since $ZX=-XZ$, and contractible loops act trivially on
the vacuum (ground states), the resultant action on the wavefunction
after braiding is $U=\exp[i\phi^{(1,0)}_{(0,1)}]=-1$.}
\end{center}
\end{figure}

Generically, surface codes appear as ground states of Hamiltonians involving
spins residing on a surface. They are best understood in terms of a coupling
graph, where each physical spin is represented by an edge on the graph.  The
Hamiltonian is a sum of two types of interactions:  a vertex term $H_v$
involving interactions between all edges that meet at a vertex, and a face
term $H_f$ describing interactions between all edges that surround a
face. An important geometric constraint is that any vertex interaction
shares at most two edges with any face interaction.
When the operators $H_v$ and $H_f$ commute, then it can be shown that the
model can support topologically ordered ground states. For example, consider
a square lattice of spin-$1/2$ particles, or qubits, with the
Hamiltonian
\begin{equation}
H=-U\sum_v H_v-J\sum_f H_f
\label{surfham}
\end{equation}
for $U,J>0$.  The interactions look like a product of operations on a cross
$H_v=\prod_{e\in +}Z_e$ or a square $H_f=\prod_{e\in\square}X_e$, where $Z$
and $X$ are the corresponding Pauli operators. Each vertex (face) operator
involves a product of two or zero $Z(X)$ operators on any face (vertex).
Since the Pauli operators anticommute, even products of them commute, hence
$[H_v,H_f]=0$.  For this reason the states can be labelled according to the
eigenvalues of the set of operators $\{H_v,H_f\}$. The ground states of $H$ are $+1$
eigenstates of this set hence the vertex and face operators are
stabilizers.  However, not all the stabilizers are
independent. For instance, on a torus, $\prod_v H_v={\bf 1}_{2^n}$ and
$\prod_f H_f={\bf 1}_{2^n}$, hence there are only $n-2$ independent
stabilizers.  Eigenspaces of $H$ are labeled according to the $\pm 1$ eigenvalues
of the independent stabilizer operators and the ground state degeneracy
is therefore $2^n/2^{n-2}=4$.  For a
surface of genus $g$ the degeneracy is $2^{2g}$ with ground
states that are indistinguishable by local observables.
 This spectral property is called
topological degeneracy.  An example of a surface code on a planar surface
with non-trivial topology is shown in Figure~\ref{fig:surf}.

The model above describes an effective $\mathbb{Z}_2$ gauge theory where the
generators of local gauge transformations are $\{H_v,H_f\}$. Excitations
above the vacuum ground state behave like particles which have anyonic
statistics. Indeed, the local operation $Z_e$ creates a $\mathbb{Z}_2$
valued charge and anti-charge with total mass $2U$ on the boundaries of edge
$e$, and $X_e$ creates a $\mathbb{Z}_2$ valued flux anti-flux pair with
total mass $2J$ on faces sharing the common boundary edge $e$. The operation
$Y_e$ creates a dyonic combination of charge anti-charge and flux anti-flux
pairs. The relative statistical obtained by winding a charge/flux dyon
$(r,s)$ around another dyon $(r',s')$ is $\phi^{(r,s)}_{(r',s')}=\pi
(rs'+sr')$ as illustrated in Figure \ref{fig:surf}. One can
generalize the Hamiltonian (Eq.~\ref{surfham}) to higher spin systems with
$d$ levels, or \emph{qudits}, that gives rise to an effective $\mathbb{Z}_d$ gauge theory with
topological degeneracy $d^{2g}$ (Bullock and
Brennen 2007).  Here the quasiparticles are $d$ valued integers and the relative statistical phase
is $\phi^{(r,s)}_{(r',s')}=\frac{2\pi}{d}
(rs'+sr')$.  As before, this phase can be derived from the commutator for
the generalized Pauli group on qudits:  $Z^rX^{s}=e^{i2\pi rs/d}X^{s}Z^{r}$

\section{Fibonacci Anyons for quantum computation}

In this section we will present probably the most celebrated non-abelian
anyonic model not only due to its simplicity and richness in structure, but
also due to its connection to the Fibonacci series. In this model there are
two different types of anyons, $0$ and $1$, that have the following fusion
rules
$$
0 \times 0 =0 ,\,\,\, 0\times 1 = 1 ,\,\,\, 1\times 1 = 0 + 1
$$
\label{fuseFib}
It is interesting to study all the possible outcomes when we fuse many
anyons of type $1$.
\begin{figure}
\begin{center}
\includegraphics[width=\columnwidth]{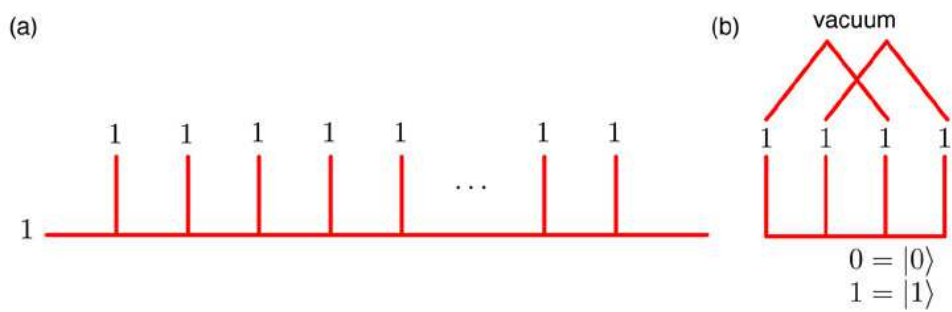}
\caption{Depiction of the fusion process for anyons.
(a)  A series of $1$ anyons are fused together going from left to right. The
first two $1$ anyons are fused and then their outcome is fused with the
next $1$ anyon and so on. (b) Four Fibonacci anyons in state $1$ created
from the vacuum can be used to encode a single logical qubit. When
restricted to the space of trivial total charge, there are two
distinguishable ways the particles can be fused together, indicated by $0$
and $1$.  This state space is sufficient to encode a logical qubit.}
\label{fig:anyonfusion}
\end{center}
\end{figure}
For this we fuse the first two anyons and then their outcome is fused with
the third $1$ anyon and the single outcome is fused with the next one and so
on. As the $1$ anyons have two possible fusion outcome states it is natural
to ask what are the number of ways, $d_1(n)$, one can fuse $n+1$ anyons of
type $1$ to yield a $1$. At the first fusing step the possible outcomes are $0$ or $1$,
giving $d_1(2)=1$. When we fuse the outcome with the next anyon then $0\times
1= 1$ and $ 1\times 1 =0+1$, resulting to two possible $1$'s coming from two
different processes and a $0$, i.e. $d_1(3)=2$. Taking the possible outcome
and fusing it with the next anyon gives a space of $1$'s which is three
dimensional $d_1(4)=3$. The series of the space dimension $d_1(n)$ when $n+1$
anyons of type $1$ are fused is actually the Fibonacci series. This
dimension, $d_1(n)$, is also called the dimension of the fusion space and is
given approximately by the following formula
$$
d_1(n) \propto \phi^n
$$
where $\phi \equiv (1+\sqrt{5})/2$ is the Golden Mean. The latter has been
used extensively by artists, such as Leonardo Da Vinci, in geometrical
representations of nature (plants, animals or humans) to describe the ratios
that are aesthetically appealing.  The above calculation is helpful to
illustrate the counting of Hilbert space dimension, however, there is a more
systematic way to compute the state space dimension using the
concept of \emph{quantum dimension}.
The quantum dimension $d_{\alpha}$ quantifies the rate of growth of Hilbert space
dimension $d_{\alpha}(n)$ when one additional particle of type $\alpha$ is added.  From another
perspective, the ratio $1/d_{\alpha}^2$ is the probability that a particle of type $\alpha$ and its antiparticle of type $\bar{\alpha}$ will annihilate.  The dimension satisfies the following product rule:
$d_ad_b=\sum_{c}N^c_{ab}d_c$.  Thus, for the Fibonacci
model the quantum dimensions for the two particles types can easily be solved for from the
fusion rules in Eq. \ref{fuseFib}: $d_0^2=d_0=1$ and $d_1^2=1+d_1\rightarrow d_1=\phi$.

This anyonic system is a good example for realization of quantum
computation. We are interested in encoding information in the fusion space
of anyons and then processing it appropriately. This is an
algorithmically equivalent, but physically completely different way of
encoding and processing information compared to the circuit model. In
the latter, qubits are positioned in space and the logical gates are applied
as a series of unitary evolutions.

There have been several proposals of
quantum computation that are conceptually different, but equivalent to
the circuit model.  By equivalent we mean that any model can simulate
another with at most a polynomial overhead in the number of operations
that determine the complexity of implementation.
One way quantum computation (Raussendorf and
Briegel 2001) starts from a large entangled state. Information is then
processed by single qubit measurements in contrast to the popular belief
that quantum computation must be reversible. Adiabatic quantum computation
is another way of processing information (Farhi \emph{ et al.} 2001). There,
the answer to the problem is encoded into the unique ground state of a
Hamiltonian. Then an adiabatic evolution is considered from a simple
starting Hamiltonian with a known ground state to the final Hamiltonian whose
ground state is a bit string encoding the answer to a problem. Topological quantum computation is yet
another ``exotic" way of encoding and processing information. The fusion
space in which information is encoded is the space of possible different
outcomes from the fusion of anyons. For the case of the Fibonacci anyons the
encoding of a qubit can be visualized by employing four $1$ anyons as in
Figure
\ref{fig:anyonfusion}(b).

\begin{figure}
\begin{center}
\includegraphics[width=\columnwidth]{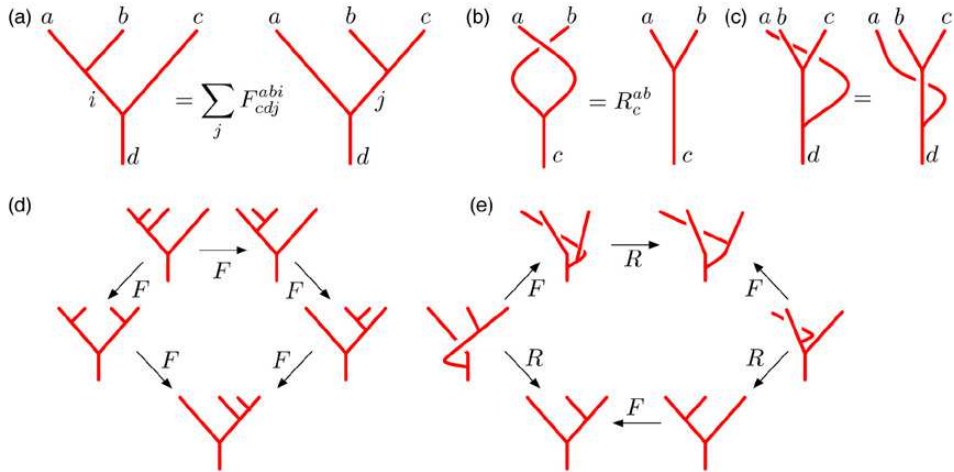}
\caption{Rules for braiding and recoupling anyons.  (a)  Recoupling for a two-vertex
interaction is obtained via a six index complex matrix $F$:  the quantum
$6-j$ symbol. (b)  A braiding operation is related to an unbraid by the $R$
matrix. (c) The intertwining operation is topologically trivial. (d)  The
pentagon identity.  Starting from a four vertex interaction, a sequence of
five recouplings returns to the original interaction. It is taken as an
axiom that this sequence is the identity mapping. (e)  The hexagon identity
relating three vertex interactions by a sequence of recoupling,
intertwining, and braiding operations.}
\label{fig:rules}
\end{center}
\end{figure}

A comprehensive set of rules for manipulating anyons is given in Figure
\ref{fig:rules}. These rules describe physically allowed operations on the
state space. For example, one can braid the anyons before fusing them.  This
operation is described by the $R$ matrix and it is depicted in Figure
\ref{fig:rules}(b). An additional evolution in the fusion space is possible by
changing the order of fusion of three anyons. For example one can fuse the
anyons $a$, $b$ and $c$ by two distinctive ways. One can first fuse $a$ with
$b$ and then with $c$ or first fuse $b$ with $c$ and then with $a$.  As
shown in Figure \ref{fig:rules}(a), these two processes are related by a six
index object $F$.  This complex valued function is the $6-j$
symbol for quantum spin networks (Kauffman 1991):
\begin{equation}
F^{abi}_{cdj}=\left\{\begin{array}{ccc}a & b & j \\c & d & i\end{array}\right\}_q.
\label{QSixJ}
\end{equation}
Analogous to angular momentum recoupling,  the $F^{abi}_{cdj}$ are physical,
i.e. non-zero, only if the triples $\{abi\}$, $\{cdi\}$,$\{adj\}$,$\{cbi\}$
are allowed products under fusion.   With the proper normalization, for each
set of labels $(abcd)$ involved in a two vertex interaction, the matrix
$F(abcd)^i_{j}$ is unitary. For the classical groups, the theory describes a
gauge theory with gauge group $G$.  For example, for $G=SU(2)$ the states
are labelled by ascending half integers $\ket{j/2}$ and the branching rules
for three indices $\{j_1 j_2 j\}$ satisfy the triangle inequality as given
by the fusion rules: $j_1\times j_2=\oplus_{j=|j_2-j_1|}^{j_1+j_2}j$.  Note
that there is no upper bound on the magnitude of angular momentum, so the
number of states on any edge is in principle infinite.  For quantum groups,
other possibilities exist.  For example, for the Chern-Simons theory with
gauge group $G=SU(2)_k$ the states are again labelled by ascending half
integers but there is an upper bound $j\leq k/2$ and the fusion rule is
modified so that $j_1+j_2+j\leq k$.  Then the quantum $6-$j symbol in
Eq. (\ref{QSixJ}) is the $q-$deformed analogue of the classical $6-$j symbol
with the value $q=e^{2\pi i/(k+2)}$.  It parameterizes the deformation of
the commutation relations for the algebra $\mathfrak{su}(2)$ viz.
$[S^z,S^{\pm}]=\pm
S^{\pm},[S^+,S^-]=\frac{q^{S^z}-q^{-S^z}}{q^{1/2}-q^{-1/2}}$.

From the pentagon rule, the following relation between the elements of the
$F$ matrices is obtained:
\begin{equation}
\sum_n F(m\ell kp)^q_{n}F(jimn)^p_{s}F(js\ell k)^n_{r}=F(jiqk)^p_{r}F(rim\ell)^q_{s}.
\end{equation}
Similarly, from the hexagon rule
\begin{equation}
R^{mk}_rF(\ell mkj)^q_rR^{m\ell}_q=\sum_p F(\ell k m j)^p_rR^{mp}_jF(m\ell k j)^q_p
\end{equation}
These two polynomial equations must be solved to obtain a consistent set of
$F$ and $R$ matrices. For example, from the fusion rules for Fibonacci
anyons, one finds the non-zero values
\begin{equation}
\begin{array}{lll}
F(1101)^1_{1}&=&F(0111)^1_{1}=F(1110)^1_{1}=F(1011)^1_{1}=1,\\
F(0000)^0_0&=&1,F(1111)=\left(\begin{array}{cc}\frac{1}{\phi} &
\frac{1}{\sqrt{\phi}} \\  \frac{1}{\sqrt{\phi}} & -\frac{1}{\phi}\end{array}\right).
\end{array}
\end{equation}
These solutions are unique up to a choice of gauge.

Inserting these values into the relations demanded by the hexagon identity, one obtains
the following $R$ matrix describing exchange of two particles:
\begin{equation}
R=\left(\begin{array}{cc}e^{4\pi i/5} & 0\\  0 & -e^{2\pi i/5} \end{array}\right).
\end{equation}
It can be shown that the unitaries $b_1=R$ and
$b_2=F(1111)RF(1111)^{-1}$ acting in the logical space $\ket{0}$ and
$\ket{1}$ are dense in $SU(2)$ in the sense that they can reproduce any
element of $SU(2)$ with accuracy $\epsilon$ in a number of operations that
scales like $O(poly(log(1/\epsilon))$ (Preskill 2004). Thus an arbitrary one qubit gate can
be performed as follows: begin from the vacuum and prepare four anyons
labelled $a_1,a_2,a_3,a_4$.  This is a subspace with total charge zero.
Braiding the first and second anyons implements $b_1$ and braiding the
second and third anyons implements $b_2$. A measurement of the outcome upon
fusing $a_1$ and $a_2$ projects onto logical $\ket{0}$ or $\ket{1}$.
Similarly, by performing braiding over $8$ anyons in state $1$, one obtains
a dense subset of $SU(d(7))$. Since $SU(4)\subset SU(13)$, we can also
implement any two logical qubit gate, e.g. the CNOT gate, with arbitrary
accuracy.  Hence the Fibonacci anyon model allows for universal computation
on $n$ logical qubits using $4n$ physical anyons (Freedman {\it et al.} 2005).


\section{A new quantum algorithm!}

The study of anyonic systems for performing quantum computation has led to
the exciting discovery of a new quantum algorithm that evaluates the Jones
polynomials (Jones 1984). These polynomials are topological invariants of
knots and links and they were first connected to topological quantum field
theories by Witten (1989). Since then they have found far reaching
applications in various areas such as in biology for DNA reconstruction or
in statistical physics (Kauffman, 1991). The best know classical algorithm for the exact
evaluation of Jones polynomials demands exponential resources
(Jaeger, Vertigan, and Welsh, 1990).  Employing
anyons only a polynomial number of resources is required to produce an
approximate answer of this problem (Freedman {\em et al.} 2003). The
techniques used by manipulating anyons resemble more an analogue computer.
Indeed, the idea is equivalent to the classical setup where a wire is
wrapped several times around a solenoid that confines magnetic flux: by
measuring the current that runs through the wire one can obtain the number
of times the wire was wrapped around the solenoid. The translation of the
corresponding anyonic evolution to a quantum algorithm was explicitly
demonstrated by Aharonov, Jones and Landau (2005).

\begin{figure}
\begin{center}
\includegraphics[width=\columnwidth]{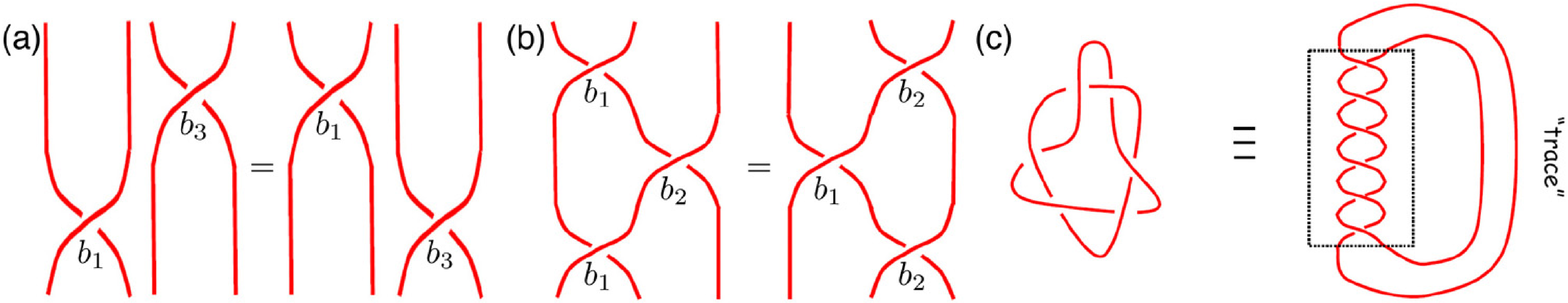}
\caption{Connection between braids and knots.  (a) Schematic representation
of the relation $b_ib_j=b_jb_i$ when $|i-j|\geq2$. Exchanging the order of
two braids does not have an effect if they are sufficiently far apart. (b)
Schematic representation of the relation $b_ib_{i+1} b_i =b_{i+1}b_ib_{i+1}$
for $1\leq i< n$. The two braidings are equivalent under simple continuous
deformations. (c) The Markov trace performed by linking the opposite ends of
the strands. It creates out of the six braidings a link between two
strands.}
\label{fig:Braiding}
\end{center}
\end{figure}

To better understand the structure of the computation, let us first
introduce a few necessary elements. The main mathematical structure behind
the evolution of anyons is the braid group $B_n$ on $n$ strands. Its
elements $b_i$ for $i=1,...,n-1$ can be viewed as braiding the world lines
of anyons. Specifically, if $n$ anyons are placed in a certain order then
the element $b_i$ describes the effect of exchanging the position of anyons
$i$ and $i+1$, e.g. in a counterclockwise fashion. Thus all possible
manipulations between the anyons can be written as a combination of the
$b_i$'s. The elements of the group $B_n$ satisfy the following relations
$$
\left\{
\begin{array}{cc}
b_i b_j =b_jb_i, & \text{for} \,\,\,|i-j|\geq 2\\
b_ib_{i+1} b_i =b_{i+1}b_ib_{i+1}, & \text{for} \,\,\, 1\leq i< n
\end{array}\right.
$$
These relations have a simple geometrical meaning, as presented in
Figure~\ref{fig:Braiding}. Note that the symmetric group $S_n$ is a
representation of $B_n$ if we impose the condition that $b_j^2=1,\,\,\forall
j$. This would be true for boson and fermions where $b_j=\pm 1$ but, as
discussed earlier, in two dimensions other possibilities exist.

The second element we need for the quantum algorithm is the introduction of
a trace that will establish the equivalence between braidings and knots or
links. A version of this tracing procedure called the Markov trace consists
of connecting the opposite endpoints of the braids together as shown in
Figure~\ref{fig:Braiding}(c). Thus, any braid with a trace gives a knot or a
link.
Surprisingly, every knot or link is equivalent to a braid with a trace as is
demonstrated by Alexander's theorem (Alexander 1923). Hence, one can
simulate a knot or a link by braiding anyons.

\begin{figure}
\begin{center}
\includegraphics[width=\columnwidth]{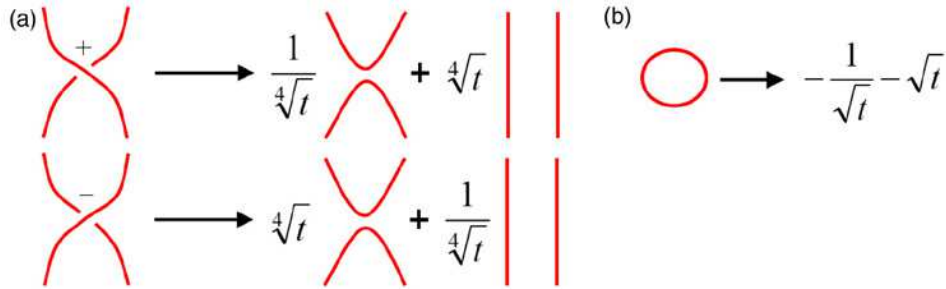}
\caption{Skein relations for disentangling knots.  (a) Each braid is
smoothed according to a linear combination of the avoiding strands
and the identity strands.  The crossing sign $\pm$ of a braid determines its
decomposition into unbraids.  (b) A closed loop contributes a complex
number.}
\label{fig:Jones}
\end{center}
\end{figure}

The general idea for relating a polynomial in a complex variable to a
certain knot or link is as follows. We first need to consider a single braid
and relate to it a linear combination of two other diagrams, as depicted in
Figure~\ref{fig:Jones}(a).
The coefficients are a function of the complex parameter $t$ that is the
variable of the Jones polynomials. This step, together with the tracing,
produces a number of closed loops that are not linked to each other. To each
closed loop the complex number $d=-\sqrt{t}^{-1} -\sqrt{t}$ is attributed,
as seen in Figure~\ref{fig:Jones}(b). These procedures are enough to
associate a Laurent polynomial with each knot or link, and subsequently to
each traced braid.

Specifically, the Jones polynomial is defined by
\begin{equation}
J_K(t)=\Big(-\frac{1}{\sqrt[4]{t}}\Big)^{-w(K)}\langle K\rangle(t)
\end{equation}
Here $w(K)$ is the writhe of the knot $K$ which is defined as the sum of the
crossing signs (as defined in figure \ref{fig:Jones}(a)) and the Kauffman
bracket of $K$ is defined by
\begin{equation}
\langle K\rangle(t)=\sum_S \langle K\ket{S}d^{L(S)-1}
\end{equation}
where $S$ is a choice of smoothings for each crossing of $K$ and $L(S)$ is
the number of loops in the state $S$. There are two choices of smoothing per
crossing so the total number of states is $2^N$ for a knot with $N$
crossings.  The product of the weights $\frac{1}{\sqrt[4]{t}}$ and
$\sqrt[4]{t}$ over all the crossings for a particular state $S$ is $\langle
K\ket{S}$.

The Jones polynomial has been shown to be a topological invariant (Jones
1984), i.e its value for a given knot $K$ is unchanged under continuous
deformations.  For two knots $K_1$ and $K_2$, if $J_{K_1}\neq J_{K_2}$ then
$K_1$ and $K_2$ are inequivalent, i.e. they cannot be mapped to each other
by continuous deformations.  Note, however, that inequivalent knots may have
the same Jones polynomial.  Computation of the Jones polynomial by a
classical computer appears to be exponentially hard due to the fact
that there are exponentially many terms to sum and no closed form for the
number of loops as a function of the resolution of the knot exists. On the
other hand it is rather easy to calculate its value with the anyons.

For illustration we sketch the quantum algorithm that evaluates the Jones
polynomial for the three strand braid group $B_3$ (see also Kauffman and
Lomonaco (2006)). First one picks a representation of the braid group. A one
dimensional unitary representation is given by $b_j=e^{i\phi},\,\,\forall j$
which describes exchange of identical particles with abelian anyonic
statistics.  The simplest non-abelian representation is given by $2\times 2$
matrices which can be parameterized as:
$\Gamma(b_j)=\frac{1}{\sqrt[4]{t}}{\bf 1}_2+\sqrt[4]{t}V_j$ where
\begin{equation}
V_1=\left(\begin{array}{cc}d & 0 \\0 & 0\end{array}\right),
\quad V_2=\left(\begin{array}{cc}d^{-1} & \sqrt{1-d^{-2}} \\
\sqrt{1-d^{-2}} & d-d^{-1}\end{array}\right),
\end{equation}
in which case
$\Gamma(b_1)\Gamma(b_2)\Gamma(b_1)=\Gamma(b_2)\Gamma(b_1)\Gamma(b_2)$. The
set $\{{\bf 1}_2,V_1,V_2,V_1V_2,V_2V_1\}$ generates what is known as a
Temperley-Lieb algebra and satisfies $V_1V_2V_1=V_1$, $V_2V_1V_2=V_2$.  The
representation becomes unitary if we choose $t=e^{-i\theta}$ for
$|\theta|\leq 2\pi/3$ or $|\theta/4+\pi|\leq\pi/6$. Any sequence of $k$
braids can be described by a braid word $r=r_1r_2\ldots r_k$ and this word
has a closure which corresponds to a knot $K_r$.   For $B_3$ we have
$r_k\in\{b_1,b_2\}$ so $\langle K_r\rangle=\langle \prod_{j=1}^k
\Gamma(r_j)\rangle=\langle (1/\sqrt[4]{t})^k {\bf 1}_2\rangle+\langle
F(r)\rangle$, where $F(r)$ is a sum of products of the matrices $V_1,V_2$.
A closed loop composed with a knot $K$ has a bracket that is $d$ times that
of $\langle K\rangle$ and the bracket of one closed loop is $1$. Hence, the
closure of the identity operation on $B_3$ represents three closed loops and
$\langle {\bf 1}_2\rangle=d^2$.  The bracket of the closure of a braid word
is a function computed by taking the trace over the carrier space of the
representation and we have $\langle
K_r\rangle(t)=(\frac{1}{\sqrt[4]{t}})^k(d^2-2)+\mbox{Tr}[\prod_{j=1}^k
\Gamma(r_j)]$.  The difficulty of computing the Jones polynomial is then
reduced to computing the trace of a product of unitaries.

Good quantum algorithms exist for computing traces of unitaries.  For
example, one can begin with a
completely mixed state of $n$ register qubits and one work qubit $w$
prepared in the pure state $(\ket{0}_w+\ket{1}_w)/\sqrt{2}$.  Applying a
sequence of controlled unitaries $\prod_{j=1}^k
\ket{1}_w{_w}\bra{1}\otimes\Gamma(r_j)$ and measuring the work qubit in the
$\hat{x}$ and $\hat{y}$ bases outputs the real and imaginary parts of the
normalized trace $\mbox{Tr}[\prod_{j=1}^k \Gamma(r_j)]/2^n$.  The
unnormalized expectation value of a unitary in a particular state $\ket{s}$
can be computing by replacing the mixed state with the pure state $\ket{s}$.
For the $U(2)$ representation of $B_3$ above, one register qubit suffices
and in fact there is no need to have a physical system with anyons. More
complicated braids over different unitary representations can be computed by
performing the controlled unitary gates using several qubits in the quantum
circuit model, or by implementing the operations by physical braiding of
anyons.
  By direct measurement of the anyons it is possible to determine the
value of the Jones polynomial. The measurement of the anyonic state can be
performed by an interference experiment as discussed below. The algorithmic
translation of this process (Aharonov, Jones and Landau 2005) provides a
quantum algorithm that can obtain the value of the Jones polynomials for
general $t$.

\section{Realizations of anyons}

\subsection{Quantum Hall Effect}

Several candidates for physical realizations of particles with anyonic
properties exist. Historically, it was first suggested that the elementary
charged excitations of a fractional quantum Hall (FQH) fluid should obey
fractional statistics (Halperin 1984, Arovas, Schrieffer and Wilczek 1984).
In the quantum Hall effect, a 2D electron gas (electron charge $q_e$ and
density $n$) moves under the influence of a magnetic field ${\bf B}$ normal
to the plane and an electric field ${\bf E}$ in the plane.  By the Lorentz
force a current is induced perpendicular to ${\bf E}$. Classically, the
resistance varies inversely with $\nu=nhc/q_e|{\bf B}|$, but in the presence
of strong magnetic fields, plateaus in the Hall resistance occur at integer, as
well at some fractional values, of the filling $\nu$. These plateaus of
quantized resistance indicate where the 2D electron gas acts as an
incompressible fluid, meaning that all charged excitations have a finite
energy gap.  Disorder plays an important role in localizing the charge
excitations, this creates a mobility gap against delocalized charged
excitations which would contribute to transport.

There is a fundamental difference in the
low energy physics between the integer and fractional filling cases.  For integer $\nu$ the gap can be understood without electron interactions because each plateau corresponds to a
completely filled Landau level and incompressibility follows by the Pauli
exclusion principle.  However, for fractional filling, the energy gap can
only be explained by including interactions, i.e. the excitations are a
collective phenomenon. Laughlin found a trial wavefunction for the FQHE at
$\nu=1/m$ for $m$ odd given by
\begin{equation}
\psi_m=\prod_{j<k}(z_j-z_k)^me^{-\sum_i |z_i|^2/4\ell_B^2}
\end{equation}
where $\{z_i\}$ are the complex electron coordinates in the plane and
$\ell_B=|c\hbar/q_eB|$ is the magnetic length (Laughlin 1983).  The
fractionalization of the charge follows by considering a quasihole
excitation at position $\zeta$ above the ground state $\psi_m$:
\begin{equation}
\phi^h(\zeta,\zeta^*)\propto \prod_i (\zeta-z_i)\psi_m
\end{equation}
The appropriate charge value is obtained by removing an electron from the
ground state $\psi_m$ producing the wavefunction $\prod_i
(\zeta-z_i)^m\psi_m$.  This state can be viewed either as a charge $e$ hole
at position $\zeta$ or as $m$ quasiholes at position $\zeta$.  Hence the
quasihole charge is $e/m$ and there is an associated qausiparticle
excitation with charge $-e/m$  (see e.g. Wen 2004).  More generically,
quasiparticles can be viewed as composite fermions in a FQH condensate at
filling $\nu=p/(2jp+1)$ for $j,p\in\mathbb{N}$ that carry charge
$q=e/(2jp+1)$ (Jain 1989).

One way to compute the fractional statistics is to wind a charge $q$
quasiparticle at position ${\bf r}_1$ around a closed path $\Gamma$ containing
another charge $q$ quasiparticle at position ${\bf r}_2$ in the $\nu$ filled
FQH state.  The Berry's phase thus accumulated is
\begin{equation}
\gamma=i\oint_{\Gamma} d{\bf r}_1 \langle \psi({\bf r}_1,{\bf r}_2)|\bigtriangledown_{{\bf r}_1}\psi({\bf r}_1,{\bf r}_2)\rangle.
\end{equation}
This quantity can be explicitly calculated by considering a time dependent potential $V({\bf r}-{\bf r}_1(t))$ which localizes particle $1$ and allows for adiabatic braiding around particle $2$.  Subtracting off the phase accumulated when no particle exists inside $\Gamma$ the
relative statistical phase is (Arovas, Schrieffer and Wilczek 1984)
\begin{equation}
\phi_{p/(2jp+1)}^{p/(2jp+1)}=\delta\gamma/2\pi=2j/(2jp+1).
\end{equation}

Theory predicts non-abelian anyons occur in FQH at special filling fractions.  This startling
discovery was originally made for the $\nu=5/2$ state by Moore and Read
(1991) and subsequent work predicted non-abelian anyons at other filling
fractions such as $\nu=12/5$ (Read and Rezayi, 1999).  In the later case,
the quasiparticles transform under a quantum group sufficiently rich to
support fault tolerant universal quantum computation (Freedman \emph{et al.}
2003). The more experimentally accessible and stable $\nu=5/2$ state is
shown in the Moore-Read model to have nonabelian composite particles.  These
anyons  have computational power that is not strictly fault tolerant but can
be made so with some amount of non-topologically protected preparation steps
(Bravyi 2006).  The braiding properties of the $\nu=5/2$ and $\nu=12/5$
quasiparticles were derived by Nayak and Wilczek (1996) and  Bais and
Slingerland (2001). While there is strong theoretical support for the
existence of non-abelian anyons in these systems, a definitive experimental
demonstration of the fact is still an active pursuit.

\begin{figure}
\begin{center}
\includegraphics[scale=0.4]{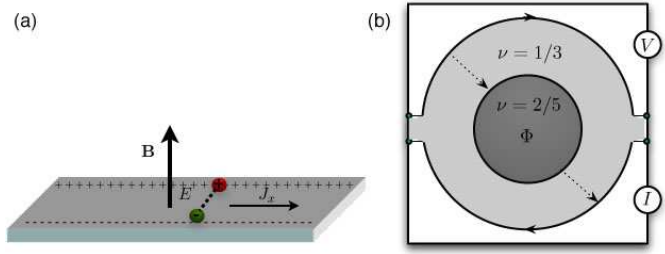}
\caption{\label{fig:qHall1}Fractional Quantum Hall fluid.
(a) Schematic of a 2D electron gas with a current ${\bf J}$ induced perpendicular to an
electric field ${\bf E}$ and a strong magnetic field ${\bf B}$. (b)
Experimental setup to observe anyonic statistics.  The system consists of a
$\nu=1/3$ fluid surrounding a $\nu=2/5$ island.  Current carrying $e/3$
quasiparticles can tunnel between inner and outer $1/3$ edges changing the
potential from $V=0$ to $V>0$.  Current along the closed path on the inner
edge gives rise to oscillations in the conductance as a function of the
enclosed flux $\Phi$ due to the Aharonov-Bohm effect.  The presence of
charge $e/5$ particles in the island affects the phase of the circulated
particles which is observed the interference pattern in the conductance.}
\end{center}
\end{figure}

\subsection{Lattice Models}

In Sec. \ref{surf} we saw that a Hamiltonian whose ground states comprise a
surface code have excited states which behave like abelian anyons.  It is
also possible to rig spin lattice models with non-abelian excitations.  In
one type of construction (Kitaev 2003; Doucot and Ioffe 2005), each spin
is endowed with a state space
whose dimension is equal to the order of a discrete non-abelian group $G$.
The vertex and face interactions can then be constructed as appropriate sums
over products of Hopf algebra elements acting on $G$ such that the
ground states are invariant under local non-abelian gauge transformations and
the excitations are massive non-abelian quasiparticles.

There are other systematic ways to construct spin lattice models
, that directly encode the
fusion rules of emergent anyons  (Turaev and O. Y. Viro 1992; Fendley
and Fradkin 2005; Freedman {\em et al.} 2004).
One such construction, known as the string net model (Levin and Wen 2005; Fidkowski \emph {et al.}, 2006), involves spins with short ranged interactions and low energy states that enforce rules for branching and reconnecting loops of string
on a background lattice.  In the model of Levin and Wen, the interactions take place amoung
qudits that reside on the oriented edges of a honeycomb lattice. Here the $d$ states of each particle correspond to the vacuum plus $d-1$ string types that enumerate the
irreducible representations of a group $G$.  The orientation of the string
relative to the edge orientation determines whether the string is of type
$s$ or its dual $s^{\ast}$.  For simplicity, we discuss only the case for
self dual strings. As with the surface code construction, the Hamiltonian
can be expressed as a sum of commuting vertex and face operators:
\begin{equation}
H=-U\sum_{v}H_v-J\sum_f H_f.
\end{equation}
Here, the vertex interaction encodes the appropriate fusion rules for the anyons
\begin{equation}
H_v=\sum_{i,j,k=0}^{d-1} \ket{\{ijk\}}\bra{\{ijk\}}
\end{equation}
where $\{ijk\}$ represents an allowed branching of the strings.  The
operator is diagonal in the spin basis $\{\ket{k}\}_{s=0}^{d-1}$ and
contains terms like $\ket{ijk}\bra{ijk}$ indicating that if the oriented
edges meet at a vertex $v$ then the allowed fusion products are $i\times
j\rightarrow k,i\times k\rightarrow j, j\times k\rightarrow i$. The face
constraints contain information about the action on the state space under
fusion of string types on the boundary of a face. It can be thought of as
the closure of Figure~\ref{fig:anyonfusion}(a) where the horizontal edges
correspond to edges on the boundary of a hexagon and the vertical edges
emanate from the hexagon, viz :
\begin{equation}
H_f=\frac{1}{\sum_s d_s^2}\sum_{s=0}^{d-1} d_sB^s_f,
\end{equation}
where $d_s$ is the quantum dimension of the particle type $s$.  The
operators $B^s_f$ are $12$-local operators that encode the gauge invariance
of the theory with appropriate weighting by the recoupling matrices
$F(ijk\ell)$.  As with the surface codes, there are a set of $N$ closed
string types which commute with the Hamiltonian $H$.  The ends of various
types of open strings correspond to quasi-particles with appropriate
statistics.

\begin{figure}
\begin{center}
\includegraphics[scale=0.4]{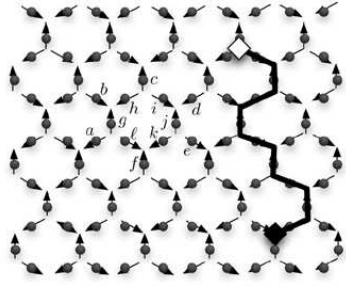}
\caption{\label{fig:spinfibbi}A spin lattice model admitting non-abelian anyons.
Here the particles reside on Kagom\'e lattice and can be identified with
oriented edges of a honeycomb lattice.  The particles each have $N$ states
enumerating the irreducible representations of a group $G$.  The states can
be thought of as labelling a string type and the Hamiltonian commutes with
closed strings of a given type.  The anyons of type $s$ exist as ends of
open strings (thick black line) of the same type.  Three body interactions
at a vertex encode the fusion rules, e.g. the particle state $\ket{a g\ell
}$ is in the span of low energy states if any pair of those indices can
combine to give the third.  The face operators encode the recoupling under
the $F$ matrices of the twelve particles around a face, e.g. $\{abcdefghijk
\ell\}$.}
\end{center}
\end{figure}

Let us consider how the spin model gives rise to Fibonacci anyons.  As
discussed above, this theory has two particle types: $1$ and $0$ with the
fusion rule $1\times 1=0 + 1$.  We recover this rule if we consider the
Chern-Simons theory $SU(2)_3$ but exclude non-integer state labels.  The
resulting theory with two particle types is also known as $SO(3)_3$.   The
two particle types are represented as string types on the edges of the
honeycomb lattice:  the vacuum (or empty particle) state $\ket{0}_e$ and
single particle state $\ket{1}_e$.  The low energy subspace is spanned by
states that satisfy the constraints

\begin{equation}
\begin{array}{lll}
H_v&=&\ket{111}\bra{111}+\ket{011}\bra{011}+\ket{101}\bra{101}+
\ket{110}\bra{110}+\ket{000}\bra{000}\\
&=&-\sum_{j=1}^3 Z_{e_j}+Z_{e_1}Z_{e_2}+
Z_{e_1}Z_{e_3}+Z_{e_2}Z_{e_3}+3
Z_{e_1}Z_{e_2}Z_{e_3}.
\end{array}
\end{equation}
Here the quantum dimensions are $d_0=1$
and $d_1=(1+\sqrt{5})/2$, and the explicit form for the face operators is
\begin{equation}
\begin{array}{lll}
B^0_f&=&\displaystyle{\sum_{r\in\{0,1\}}} \ket{rrrrrr}\bra{rrrrrr}
\otimes \ket{rrrrrr}{_{\hexagon}}
{_{\hexagon}\bra{rrrrrr}}\\
B^1_f&=&\displaystyle{\sum_{a,b,c,d,e,f\in\{0,1\}}}
\ket{abcdef}\bra{abcdef}\otimes\displaystyle{\sum_{\substack{g,h,i,j,k,\ell,g',h',
\\ i',j',k',\ell '\in\{0,1\}}}}\ket{g'h'i'j'k'\ell '}{_{\hexagon}}
{_{\hexagon}\bra{ghijk\ell}}\\
&&F(a\ell1g')^{g}_{\ell'}
F(bg1h')^h_{g'}F(ch1i')^i_{h'}F(di1j')^j_{i'}F(ej1k')^k_{j'}
F(fk1\ell ')^{\ell}_{k'}\\
&&
\end{array}
\end{equation}
where the $F$ matrices are as derived above. Notice that $B^s_f$ is block
diagonal in the basis of the edge qubits that eminate from the hexagon (see
Figure \ref{fig:spinfibbi}). The operator of $B^s_f$ can be obtained
constructively by deriving the transition matrix between one string
configuration around a face that is fused with a virtual $s$ type string
inside to produce another valid string configuration around the face.

The string net picture is a beautiful illustration of a microscopic model
that realizes topological phases.  A major stumbling block is that it
requires a Hamiltonian that includes $12$-body interaction terms, i.e. it is
$12$-local, making a physical realization of such a model implausible.
Nature tends to favor binary interactions and it would be helpful to have a
microscopic spin model with emergent anyons that involved $2$-local
interactions only.  In such a case, one could imagine engineering
interactions in the laboratory using fields to coherently couple quantum
mechanical spins in the desired manner.  Yet most spin lattice models with emegent anyons
require at least $4$-local interactions on a 2D lattice.\footnote{There does exist
a $2$-local Hubbard model on a Kagom\'e lattice with a topological phase that
is universal for quantum computation but the interactions are highly anisotropic
in spin and space (Freedman \emph{et al..} 2005)}.  How is one to
achieve a quantum simulation of such models?  One possible resolution is to obtain the
$k$-local interaction as an effective model in perturbation theory. Another
possibility is to use ancillary particles to mediate the $k$-local
interaction (Oliveira and Terhal 2004). The caveats to both these approaches
are: first, the effective Hamiltonians in the ground states that approximate
the model Hamiltonian occur only at high order in perturbation theory and
hence have a small coupling strength or require energy gaps in the ancillae
which scale with the system size, and second, the microscopic interactions
are highly anisotropic in spatial and spin degrees of freedom.

\begin{figure}
\begin{center}
\includegraphics[width=\columnwidth]{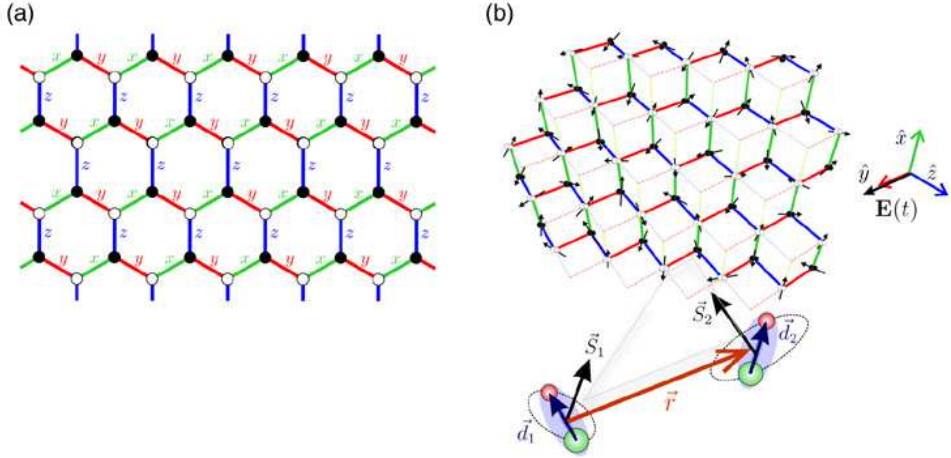}
\caption{\label{fig:AMOimp1}A honeycomb spin lattice model due to Kitaev which
has anyonic excitations. (a) The coupling graph depicted on a plane.  The
lattice is bipartite with physical qubits indicated by black and white dots
and the color of the links indicating the anisotropy of the interaction.
(b)  Proposed realization using polar molecules trapped in an optical
lattice.  The spin corresponds to a single valence electron of an
alkaline earth monohalides.  The molecules are prepared in rovibrational
ground states and trapped one molecule per lattice well with nearest
neighbor spacing $\lambda/2$ where $\lambda$ is an optical wavelength.   The
molecules are made to interact using microwave fields ${\bf E}(t)$ which
induce dipole-dipole interactions.  For a judicious choice of microwave
frequencies, polarizations, and intensities, the effective interaction in
the ground states between two molecules separated by ${\bf r}=r\hat{r}$ is
$C(r)\vec{\sigma}\cdot \hat{r}\otimes \vec{\sigma}\cdot \hat{r}$, where
$C(r)$ is a function that is large near microwave resonance at $r=\lambda/2$
but decays rapidly beyond this separation.   The appropriate terms in
$H_{\rm hc}$ are achieved when nearest neighbors form orthogonal triads
along $\hat{x},\hat{y},\hat{z}$ in real space.  This is engineered by
trapping the molecules in two staggered triangular optical lattices such
that the nearest neighbor coupling graph is a honeycomb lattice.}
\end{center}
\end{figure}

One $2$-local spin model that does have anyonic excitations is the following
anisotropic interaction on a honeycomb lattice (see
Figure~\ref{fig:AMOimp1})
\begin{equation}
H_{\rm hc}=-J_x\sum_{x-{\rm links}}X_iX_j-J_y\sum_{y-{\rm links}}
Y_iY_j-J_z\sum_{z-{\rm links}}Z_iZ_j.
\end{equation}
This model is exactly solvable (Kitaev 2006; Pachos 2006) and has two
distinct phases. When the couplings satisfy the triangle inequalities
$|J_x|\leq |J_y|+|J_z|, |J_y|\leq |J_x|+|J_z|,  |J_z|\leq |J_x|+|J_y|$ then
the system is gapless, but it becomes gapped in the presence of a magnetic
field and has non-abelian anyonic excitations. Otherwise, the system is
gapped and for the case where one of the couplings is much greater than the
others, the excitations are described by abelian anyons in a $\mathbb{Z}_2$
gauge theory. Recently, two physical constructions of this model were
proposed using atomic (Duan, Demler and Lukin 2003) and molecular arrays
(Micheli, Brennen, and Zoller 2006).  Both proposals involving trapping of
the particles, one per well, in an optical lattice which is a periodic
potential produced by interfering standing waves of laser light. Spin is
encoded in internal hyperfine states and the particles interact either by
nearest neighbor spin dependent collisions or by field induced dipole-dipole
interactions (see Figure \ref{fig:AMOimp1}).

\section{Observation of anyons}
\label{Obs}

The signature of anyonic properties is their nontrivial evolution under
braiding and this behavior can be probed via interference measurements.
Quite recently, an interferometer type experiment with quasiparticles
in a FQH fluid found evidence supporting abelian anyonic statistics
 (Camino, Zhou and Goldman 2005).  In
that experiment, a quasiparticle with charge $e/3$ in a $\nu=1/3$ FQH fluid
makes a closed path trajectory around an island of $\nu=2/5$ FQH fluid (see
Figure~\ref{fig:qHall1}b) and the relative statistics are probed.
Interference fringes manifest as peaks in the conductance as a function of
the magnetic flux through the $\nu=2/5$ island.  The experiment observed
periodic modulation consistent with the excitation of ten $q=e/5$
quasiparticles in the island fluid.  Adapting the Berry's phase argument for
encircling several quasiparticles (Jain 1993), this evidence then implies a
relative statistics of $\phi_{2/5}^{1/3}=-1/15$ when a charge $e/3$
quasiparticle encircles one $e/5$ quasiparticle of the $\nu=2/5$ filled
fluid. Nevertheless, the unambiguous detection of anyons is still considered
an open issue (Kim, 2006; Rosenow and Halperin 2007).

Several new theoretical proposals have been made for observing non-abelian
statistics in FQH states (see e.g. Das Sarma, Freedman, and Nayak 2005; Stern and Halperin 2006; Bonderson,
Kitaev and Shtengel 2006; Bonderson, Shtengel, and Slingerland 2007).
Ironically, experimental signature of the non-abelian anyons may be more
robust than the abelian case. Indeed, in the abelian case the braiding
results in phase factors that are both geometrical and dynamical in nature
and thus hard to distinguish. To the contrary, the non-abelian anyons cause
a change in the amplitude of the participating states which is easily
distinguished from spurious dynamical phase factors.

\begin{figure}
\begin{center}
\includegraphics[scale=0.35]{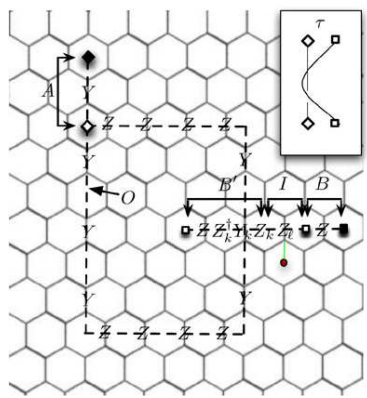}
\caption{\label{fig:intfr1}Interferometer for measuring statistics of abelian anyons.
The system is the gapped phase of the honeycomb mode $H_{\rm hc}$.  The
inset illustrates the braiding of a flux around a charge.  The steps are as
follows:  (1)  beginning in the vacuum state $\ket{\lambda_g}$ prepare a
charge anti-charge pair in region $A$ and a flux anti-flux pair in region
$B$.  (2) Apply the ``beam splitter" operation $e^{-i\frac{\pi}{4}Z_{\ell}}$
in region $I$; this creates a delocalized state of the flux $\square$. (3)
Adiabatically drag the flux particle left along the trajectory $B'$.  (4)
Adiabatically drag the charge particle $\diamond$ counterclockwise around
the flux along the path $O$. (5) Adiabatically drag the flux right along
$B'$.  (5) Apply the inverse beam splitter operation
$e^{i\frac{\pi}{4}Z_{\ell}}$. (6) Measure the location of the flux, i.e.
measure $\langle Z_{\ell}\rangle=\sin(\beta+\phi^1_1)$.  Here $\beta$ is the
accumulated dynamical and Berry's phases and $\phi^1_1$ is the mutual
statistics of the charge and flux.  (7)  Repeat the above steps in the order
$(1,2,4,3,5,6)$ such that the braid is trivial and $\langle
Z_{\ell}\rangle=\sin(\beta)$.  It is then possible to extract  $\phi^1_1$
which is equal to $\pi$.}
\end{center}
\end{figure}

It should also be possible to observe anyonic statistics in spin lattice
realization. As an example consider the honeycomb model $H_{\rm hc}$ in the
regime where $|J_z|\gg |J_x|,|J_y|$. In the low energy sector, each pair of
spins $e_1,e_2$ on a $z-$link gets mapped to an effective two level system.
The $\mathfrak{su}(2)$ algebra for each spin is spanned by the two body
operators $\tilde Z_e=Z_{e_1}$, $\tilde X_e=X_{e_1}X_{e_2}$, $\tilde
Y_e=Y_{e_1}X_{e_2}$. In perturbation theory to fourth order in
$|J_{\perp}/J_z|$, the effective Hamiltonian in the ground states is then
(Kitaev 2003)
\begin{equation}
H_{\rm eff}=-J_{\rm eff}\sum_{\diamond}\tilde Y_{\rm left}\tilde Z_{\rm
up}\tilde Y_{\rm right}\tilde Z_{\rm down}
\end{equation}
where $J_{\rm eff}=J_x^2J_y^2/16|J_z|^3$ and the operator subscripts
indicate the location of the $z-$links which are at the corners of a
diamond.  This model was first introduced by Wen (2003) and is
locally equivalent to the surface code Hamiltonian of Eq. (\ref{surfham}).
Here flux anti-flux pairs are created on the left and right of a $z-$link
$e$ by applying $Z_e$ to the vacuum.  Charge anti-charge pairs are created
above and below $e$ by applying $Y_e$.  Dyonic combinations are created by
applying $X_e$.  The mass of each particle is $J_{\rm eff}$.  In Figure
\ref{fig:intfr1} we sketch an interferometer type experiment to extract the
mutual statistics of anyonic charge and flux.  The particles are braided
adiabatically along non-contractible loops. Adiabatic motion could be
achieved by modifying the Hamiltonian such that $H'(t)=H_{\rm hc}+\sum_{e\in
Path}\delta J_e(t) Z_{e_1}Z_{e_2}+\kappa(t)Z_e(t)$.  For a properly tuned
$\delta J_e(t)$ this creates a reduced effective mass for the particle along
the trajectory and hence makes it energetically favorable to follow the
intended path. Such a protocol could be implemented with atoms or molecules
trapped in an optical lattice provided one had single lattice site
addressability.  This can be accomplished in principle with gradient field
spectroscopy using lasers with shaped intensity profiles (Zhang, Rolston and
Das Sarma 2006).


\section{Criteria for TQO and TQC}

Clearly there is need for many technological advances in order to realize
experimentally topological order and topological quantum computation.
Topological order corresponds to the property of long range correlations (or
long range coherence). This is best captured by the concept of topological
entropy (Hamma, Ionicioiu, and Zanardi 2005; Kitaev and Preskill  2006; Levin and Wen  2006) given by $S_{\rm{topo}}=\log D$ where $D=\sqrt{\sum_j d_j^2}$ is the total quantum dimension. The latter is non-zero only when there exists a loop
operator defined on the two dimensional system that has a non-zero
expectation value when the size of the loop is arbitrarily increased.
Note, however, that there is some debate about whether topological
entropy is a necessary and sufficient criterion for topological order
(Nussinov and Ortiz, 2007).  For the case of topological quantum
computation, one should be able to bring a system in such a phase, then
create anyons, braid them and finally measure them. This is a hard task and
it is still not clear which technology will serve best to perform all these
steps. For that we would like to introduce a list of criteria a physical
system has to satisfy to be able to support TQO and which supplementary ones
we have to use to be able to perform TQC.

Here is a list of criteria a system has to satisfy in order to be able to
support TQO. In particular we will focus on a lattice system of qubits or
qudits defined in two dimensions. Keep in mind that, to a large degree, the
required steps depend on the particular topological theory one wants to
realize and on the particular employed physical system.
\begin{itemize}
\item Initialization (Creation of highly entangled state)
\item Addressability of qubits (Anyon generation and manipulation)
\item Measurement (Entropic study of the ground state or interference of
anyons)
\item Scalability (Large systems)
\item Low decoherence (Protected encoding subspace)
\end{itemize}
These criteria are required for the following reason:
\begin{itemize}
\item It should be possible to create a highly entangled state between all of
the qubits of the system that has $S_{\rm{topo}}\neq 0$. It can be created
either by a coherent process (dynamical preparation, adiabatic evolution) or
a dissipative process such as a cooling mechanism.
\item To create and manipulate anyons one has to be able to address
the qubits of the system and perform local, possibly multiqubit, operations.
Creation of anyons is a classical process and should be possible to happen
in finite time.
\item One should be able to measure the type (or color) of anyons
by a suitable interference process.  There are several ways this might be
done including but not limited to:  guiding individual quasiparticles along braiding paths as
described in Sec. \ref{Obs}, weaving global string operators to measure the action on degenerate ground states, and braiding large defects (holes) followed by a measurement of the outcome under fusion.
\item The system has to be sufficient large to be able to support the
presence of several anyons and to perform braidings between them.
\item The system should be sufficiently isolated from the environment
or the temperature should be low enough compared to the energy gap so that
the topological properties of the ground state or the statistical properties
of the anyons can be read.

\end{itemize}
Requirements for protected topological phase and computation. These strongly
depend on the employed physical system.
\begin{itemize}
\item The presence of a Hamiltonian that has the highly entangled state as
its ground state and creates an energy gap above it that protects it from
small perturbations.
\item Trapping of anyons with local potentials (caused with external
classical fields) that facilitate their generations at a certain point.
Subsequently anyons should be able to be moved adiabatically to perform
braiding (only one of them is necessary). If anyons move {\em outside} of
trapping potentials then decoherence (error) may occur.
\item Addressability of the different energy levels may be necessary for
identifying anyons.
\end{itemize}

We emphasize that the above set of criteria is a helpful guide but may
prove to be overly restrictive for a more general class of models with topological
order.  For example, it has been found that universal quantum computation is possible
using only ground state manipulation, i.e. without quasiparticle braiding, using
brane-net condensates in three dimensional spin lattices (Bombin and Martin-Delgado 2006).  Furthermore, the possibility to perform global operations on finite sized systems may obviate
the need to address excited states.

\section{Conclusions}

Topological quantum computation offers a rich arena for exciting
developments in theoretical and experimental physics. It has created a new
playground with unique links between information theory, physics and
mathematics. Research in this area has proven already fruitful for quantum
information and fundamental sciences and we expect to witness exciting
developments in the near future. In this article we have reviewed some of
the fascinating recent developments in this field, emphasizing the fundamental connection
between physics and topology.  An in depth study of the involved concepts and
techniques can be found in the provided bibliography and in references
therein.

\end{document}